\documentstyle[11pt,aaspp4,tighten,flushrt]{article}

\newcommand{\kms}{\mbox{${\rm\,km\,s}^{-1}$}}

\newcommand{\kpc}{\mbox{$\rm\,kpc$}}
\newcommand{\mpc}{\mbox{$\rm\,Mpc$}}
\newcommand{\gyr}{\mbox{$\rm\, Gyr$}}
\newcommand{\etal}{{\it et al.\/}}
\newcommand{\cf}{{\it cf.\/}}
\newcommand{\eg}{{\it e.g.,\/}}
\newcommand{\ie}{{\it i.e.,\/}}
\newcommand{\beq}{\begin{equation}}
\newcommand{\eeq}{\end{equation}}

\lefthead{Statler, Dejonghe, \& Smecker-Hane}
\righthead{Mass Distribution in NGC 1700}

\begin{document}

\title{The Three-Dimensional Mass Distribution in NGC 1700\footnotemark}

\footnotetext{Observations reported in this paper were obtained at the
Multiple Mirror Telescope Observatory, a joint facility of the University
of Arizona and the Smithsonian Institution.}

\author{Thomas S. Statler}
\affil{Department of Physics and Astronomy, Ohio University, Athens, OH 45701}
\author{Herwig Dejonghe}
\affil{Sterrenkundig Observatorium, Universiteit Gent, Krijgslaan 281, B-9000
Gent, Belgium}
\and
\author{Tammy Smecker-Hane}
\affil{Department of Physics, University of California, Irvine, CA 92717}

\vskip -3.6in {\hfill \sl To Appear in The Astronomical Journal, 1999 January}
\vskip 3.5in

\begin{abstract}

A variety of modeling techniques is used with surface photometry from the
literature and recently acquired high-accuracy stellar kinematic data to
constrain the three-dimensional mass distribution in the luminous cuspy
elliptical galaxy NGC 1700. First, we model the radial velocity field and
photometry, and, using a Bayesian technique, estimate the triaxiality $T$
and short-to-long axis ratio $c$ in five concentric annuli between
approximately $1$ and $3$ effective radii. The results are completely consistent
with $T$ being constant inside about $2.5 r_e$ ($36\arcsec$; $6.7 h^{-1} \kpc$).
Adding an {\em assumption\/} of constant $T$ as prior information gives
an upper limit of $T < 0.16$ (95\% confidence); this relaxes to $T < 0.22$
if it is also assumed that there is perfect alignment between the angular
momentum and the galaxy's intrinsic short axis. Near axisymmetry permits
us then to use axisymmetric models to constrain the radial mass profile.
Using the Jeans (moment) equations, we demonstrate that 2-integral,
constant-$M/L$ models cannot fit the data; but a 2-integral model in which
the cumulative enclosed $M/L$ increases by a factor of $\sim 2$ from the
center out to $12 h^{-1} \kpc$ can. Three-integral models constructed by
quadratic programming show that, in fact, {\em no\/} constant-$M/L$ model
is consistent with the kinematics. Anisotropic 3-integral models with
variable $M/L$, while not uniquely establishing a minimum acceptable halo
mass, imply, as do the moment models, a cumulative $M/L_B \approx 10 h$
at $12 h^{-1} \kpc$. We conclude that NGC 1700 represents the best stellar
dynamical evidence to date for dark matter in elliptical galaxies.

\end{abstract}


\section{Introduction}

Evidence is accumulating for a dichotomy between two classes of elliptical
galaxies: high-luminosity, slowly rotating systems with shallow central
brightness cusps and a tendency for boxy isophotes, and lower luminosity,
rotationally supported systems with steeper central cusps and a tendency
for diskiness (Lauer \etal\ \markcite{Lau95}1995, Kormendy \& Bender
\markcite{KoB96}1996,
Faber \etal\ \markcite{Fab97}1997, and references therein). Photometric
shape differences between high- and low-luminosity ellipticals (Tremblay
\& Merritt \markcite{TrM96}1996) are roughly consistent with
the general expectation that rapidly rotating systems should have oblate
axisymmetry while slow rotators are more likely triaxial.
There are exciting suggestions that central cusps may drive
evolution towards axisymmetry through orbital stochasticity (Merritt \&
Fridman \markcite{MeF96}1996, Merritt \& Valluri \markcite{MeV96}1996).
But where the causal arrows actually point among these various
properties is still far from clear.

There is consequently a growing need for intrinsic shape measurements for
a subsample of well studied ellipticals. This paper continues our efforts
to obtain such measurements. Previously we have derived approximate
shapes for the ``standard'' elliptical NGC 3379 (Statler
\markcite{Sta94c}1994c; hereafter S94c) and NGC 4589 (Statler
\markcite{Sta94b}1994b; S94b)
using photometry and kinematic data from the literature. In this paper we
take up the case of the high surface brightness elliptical NGC 1700. The
availability of high quality multi-position-angle kinematics allows an
intrinsic shape determination of unprecedented precision, as well as
interesting constraints on the presence of dark matter in the outer
parts of the system.

NGC 1700 is an elliptical galaxy of fairly average luminosity
($M_B=-22.3$ for $H_0=50\kms$ $\mpc^{-1}$), but comparatively high central
dispersion ($\sim 230 \kms$) and small ($14\arcsec$) effective radius. This
puts it roughly $2\sigma$ off the Fundamental Plane in the sense
of having uncharacteristically low $M/L$ for objects of this mass. NGC
1700 is notable for being the most luminous of the steep-cusped ``power law''
ellipticals in the sample observed by HST (Lauer \etal\
\markcite{Lau95}1995; most recent compilation in Faber \etal\
\markcite{Fab97}1997).
The high surface brightness at $r_e$ makes it particularly attractive
for a study of intrinsic shape and $M/L$ gradient, since long-slit
spectra with good signal-to-noise can be obtained out to several optical
scale lengths.

Statler, Smecker-Hane, \& Cecil \markcite{SSC}(1996, hereafter SSC)
present kinematic profiles for NGC 1700 out to $\sim4r_e$ on four
position angles. This is the first data set to meet the
requirements for accurate shape fitting spelled out in \markcite{Sta94b}S94b;
accuracy in the mean velocity field (VF) is better than $10\kms$ inside
$2r_e$ and better than 15\% of the maximum inside $3r_e$. In addition to
the usual profiles of $v$, $\sigma$, $h_3$, and $h_4$, \markcite{SSC}SSC
derive profiles of the velocity field asymmetry parameters $(V_1,V_2,V_3)$
(Statler \& Fry \markcite{StF94}1994), and show all three to be consistent
with zero over a factor $\sim 10$ in radius. From the VF's
high degree of symmetry, \markcite{SSC}SSC surmise that the triaxiality
of NGC 1700 is probably small; {\em i.e.,\/} that the shape is close to
oblate. They also argue qualitatively that the shallow slope of the
projected RMS velocity profile is consistent with a standard dark halo. Our
objective in this paper is to make both of these statements quantitative.

We proceed as follows: in Section 2 we apply the dynamical shape fitting
method of \markcite{Sta94b}S94b to \markcite{SSC}SSC's kinematic data
and photometry from Franx \etal\ \markcite{FIH89}(1989). We find that
NGC 1700 is very nearly oblate. This result lets us
exploit existing tools for modeling axisymmetric systems to assess the
mass-to-light ratio. In \S\ 3 we demonstrate that two-integral models
with constant $M/L$ are strongly inconsistent with the data, while a
model with $M/L$ increasing outward by a factor of $\sim 2$ over the
range of the observations is viable. We then construct anisotropic
three-integral models, and show that even anisotropy cannot completely
obviate the need for an $M/L$ gradient. We summarize and discuss the broader
implications of these results in \S\ 4.

\section{Intrinsic Shape\label{s:shape}}

\subsection{Method\label{s:shapemethod}}

The shape fitting method is described in detail in \markcite{Sta94b}S94b
and \markcite{Sta94c}S94c. The underlying dynamical models (Statler
\markcite{Sta94a}1994a, hereafter S94a) rest on the following assumptions:
(1) mean rotation in the galaxy arises from internal streaming in a
potential with negligible figure rotation; (2) the mean-motion streamlines
of short-axis and long-axis tube orbits can be represented by coordinate
lines in a confocal coordinate system;\footnote{Anderson \& Statler
\markcite{AnS98}(1998) demonstrate the validity of this assumption for
realistic potentials and show that the triaxiality parameter recovered
from the streamlines closely agrees with the triaxiality of the mass
distribution.} (3) the luminosity density $\rho_L$ is stratified on similar
ellipsoids, {\em i.e.,} $\rho_L (r,\theta,\phi) = \bar{\rho}_L(r)
\rho_L^\ast(\theta,\phi)$; and (4) the velocity field obeys a ``similar
flow'' ansatz outside the tangent point for a given line of sight,
${\bf v}(r,\theta,\phi) = \bar{v}(r) {\bf v}^\ast(\theta,\phi)$.

Assumptions (3) and (4) are used only for
projecting the models. As long as $\bar{\rho}_L(r)$ and
$\bar{\rho}_L(r) \bar{v}(r)$ decrease faster than $r^{-2}$, nearly all
of the contribution to the projection integrals will
come from radii near the tangent point, and the result will be
insensitive to the structure at larger $r$. We
refer to this approximation as quasi-local projection.
The requirement on $\bar{\rho}_L(r) \bar{v}(r)$ generally limits the
validity of the models to regions where the rotation curve
is not steeply rising, which for NGC 1700 means $r \gtrsim r_e$.

A single model is described by the parameters $(T,c_L,\Omega,{\bf d})$,
where $T$ is the triaxiality of the total mass distribution, $c_L$ is the
short-to-long axis ratio of the luminosity distribution, $\Omega \equiv
(\theta_E,\phi_E)$ is the orientation of the galaxy relative to the line of
sight, and the vector ${\bf d}$ represents the remaining dynamical
parameters which are defined in S94b. The projected model predicts
the ellipticity and the radial velocities on each sampled position angle,
from which we calculate the likelihood of the
observations. Repeating for $\sim 10^7$ models covering
the full parameter space, we determine the multidimensional
likelihood $L(T,c_L,\Omega,{\bf d})$. The Bayesian
estimate of the galaxy's shape is then the probability distribution
$P(T,c_L)$, obtained by multiplying the likelihood by a
model for the parent distribution
from which Nature drew the galaxy and integrating over all other
parameters:
\beq
\label{e:margtwod}
P(T,c_L) = \int d\Omega\, d{\bf d}\, F_p^\ast(T,c_L,\Omega,{\bf d})
L(T,c_L,\Omega,{\bf d}).
\eeq
In Bayesian terminology, the integrand is called the {\em posterior
density\/} and $F_p^\ast$ the {\em prior}. The {\em marginal posterior
density\/} $P$ is the integral of the posterior over the
{\em nuisance parameters}. We use $P$ to stand for any marginal density;
for instance, the joint distribution in shape and orientation is given by
\beq
\label{e:margfourd}
P(T,c_L,\Omega) = \int d{\bf d}\, F_p^\ast(T,c_L,\Omega,{\bf d})
L(T,c_L,\Omega,{\bf d}).
\eeq
We can also factor the parent distribution according to
\beq
F_p^\ast(T,c_L,\Omega,{\bf d}) = {1 \over 4 \pi} F_p(T,c_L)
F_p^d({\bf d}).
\eeq
The first factor describes a population with random
orientations,\footnote{Actually, 
the population of ellipticals may have a preferred orientation if, for
instance, a significant number of E's are misclassified S0's seen far
from edge-on.} and the dependence
on ${\bf d}$ can be factored out without loss of generality because the
${\bf d}$ themselves can implicitly be functions of $T$ and $c_L$.

In modeling NGC 3379 and NGC 4589 it was necessary to average the kinematics
over a single large radial bin on each PA, yielding only a single
average shape for each system. For NGC 1700 we can extend the method
because we have excellent data over nearly a factor of 4 in
radius in the region consistent with the assumptions of the models. 
SSC's radial binning gives 5 independent measurements along
each PA, with bin centers at $12\farcs6$, $15\farcs6$, $21\farcs0$,
$30\farcs6$, and $46\farcs8$. We first model each
annulus independently in the quasi-local approximation. Then, assuming
that the intrinsic principal axes remain aligned through the galaxy,
we can combine the posterior probabilities requiring that the five
annuli match their local kinematics and ellipticities in the {\em same\/}
orientation. If we further assume that the triaxialities of the mass and
light distributions are equal, we can add the constraint that the
triaxiality profile must reproduce the observed isophotal twist.
This procedure is described more fully in Appendix A.

We argued previously (SSC) that the parallelogram-shaped isophotal distortions
seen beyond $R\approx 60\arcsec$ arise from an incompletely phase-mixed,
differentially precessing ring or disk, presumably acquired by accretion.
If so, then the photometric and kinematic twists that start
near $R\approx 40\arcsec$ may be the result of intrinsic twisting of the
isodensity surfaces rather than a shape gradient in an intrinsically
aligned system. In fact, we strongly suspect that this is the case;
nonetheless, in this paper we proceed under the assumption that the isodensity
surfaces are aligned and derive constraints on the shape gradient in that
context. If the outermost annulus is actually misaligned, our results for
the inner four annuli still hold and are virtually unaffected, as we
have verified by recomputing the posterior densities excluding the last annulus.

\subsection{Data}

\subsubsection{Photometry}

{\footnotesize
\begin{deluxetable}{ccccccc}
\tablewidth{0pc}
\tablecaption{Photometric and Kinematic Data Used in Shape Fitting
\label{t:data}}
\tablehead{\colhead{$\arcsec$} & \colhead{$\epsilon$} &
\colhead{PA} & \colhead{$ v_0 $} &
\colhead{$ v_{225} $} & \colhead{$ v_{270} $} &
\colhead{$ v_{315} $}}
\startdata
$12.6$ & $0.262 \pm .006$ & $89.5 \pm 1.0$ &
   $\phantom{1}{-5.6} \pm 13.1$ & $57.7 \pm \phantom{1}5.5$ & $\phantom{1}94.8 \pm \phantom{1}6.0$ & $54.9 \pm \phantom{1}8.4$ \nl
$15.6$ & $0.270 \pm .010$ & $89.0 \pm 1.0$ & 
   $\phantom{1}{-4.6} \pm \phantom{1}6.5$ & $73.5 \pm \phantom{1}5.6$ & $104.0 \pm \phantom{1}6.1$ & $71.8 \pm 10.4$ \nl
$21.0$ & $0.283 \pm .004$ & $89.0 \pm 1.0$ & 
   $\phantom{-}12.5 \pm \phantom{1}9.7$ & $79.2 \pm \phantom{1}6.9$ & $113.8 \pm \phantom{1}6.8$ & $71.3 \pm \phantom{1}7.1$ \nl
$30.6$ & $0.292 \pm .003$ & $89.8 \pm 1.0$ & 
   $\phantom{1}{-9.7} \pm 11.9$ & $77.9 \pm 10.3$ & $113.0 \pm \phantom{1}9.8$ & $80.3 \pm 10.8$ \nl
$46.8$ & $0.296 \pm .010$ & $95.8 \pm 3.2$ & 
   $-10.6 \pm 30.3$ & $53.1 \pm 26.9$ & $133.8 \pm 15.0$ & $94.6 \pm 17.4$ \nl
\enddata
\end{deluxetable}
}

Ellipticity and position angle profiles are taken from Franx \etal\ 
\markcite{FIH89}(1989) and averaged over the approximate
ranges of SSC's radial bins. At smaller radii the intervals over which
the profiles are averaged overlap slightly in order to
better reflect the ellipticity gradient inside $25 \arcsec$; this has a
negligible effect on the adopted profile but (intentionally)
increases the error bars
where the gradient is steeper. The results are given in columns 2 and 3 of
Table \ref{t:data}. Data are weighted according to the inverse
square of Franx {\em et al.}'s error bars. The tabulated uncertainty
is the larger of the mean observational error per data point and the variance
over the bin.

The logarithmic slope of the surface brightness profile wavers
non-monotonically between about $-1.7$ and $-2.3$ for $10\arcsec \lesssim r
\lesssim 55 \arcsec$.
For simplicity in projection we adopt a power-law radial profile
for the luminosity density, $\bar{\rho} \sim r^{-k}$, compute models for
$k = (\case{11}{4}, 3, \case{13}{4})$,
and average the results. It turns out that the results
are insensitive to the choice of $k$.

\subsubsection{Kinematics}

Mean velocities, corrected for the non-Gaussian shape of the LOSVD, are
taken directly from columns 10 and 11 of SSC's Table 2 (as plotted in
their Figure 10), folded about the origin, and averaged. Results are given in
columns 4 -- 7 of Table \ref{t:data} for PAs 0 (near minor axis), 225,
270, and 315.

The kinematic sampling is at a different position angle with respect to the
photometric axes in each annulus because the major axis PA is not strictly
constant in radius. However, the PA twist for $10\arcsec < r < 35\arcsec$ is
so small that we assume the sampling in the inner four annuli is the
same for the purpose of calculating the projected model velocities;
this saves a factor of 3 in computing time.

Over the region being modeled, the observed mean rotation amplitude is 
constant or slowly
increasing with radius, though on some PAs the rotation curve does fall
at large $R$. Again for simplicity in projection, we take a power-law form
for the velocity scaling law $\bar{v}(r) \sim r^{-l}$ and compute models
for $l=(0,\pm \case{1}{2})$. As for the density index $k$, the results are
not sensitive to the value of $l$, and we simply average the derived
probability densities.

\subsubsection{Forbidden Velocities on PA 225?\label{s:forbidden}}

SSC reported a striking reversal in the rotation profile along PA 225, at
$r \gtrsim 40\arcsec$ to the northeast,
and argued that this feature, most prominent in the
outermost data point, is real. Two of us (TSS and TS-H) subsequently
obtained additional spectra along this PA with the MMT Red Channel on
1995 February 3--4 UT. The instrument was rotated $180\arcdeg$ relative
to the earlier setup, and the galaxy was displaced $\sim 60\arcsec$ to
the southwest, putting the region of interest near the center of the slit.
Reductions were performed as described by SSC.

The resulting rotation curve shows no velocity reversal. We now believe
the earlier result to have been spurious, caused by a distortion of the
cross-correlation peak by some unnoticed or inadequately corrected systematic
effect (possibly scattered light) in the weakly exposed outer part of the
image.

The data were reduced, unfortunately, after the present models had all
been computed, but the results for intrinsic shape are not compromised.
The radius at which the velocity had appeared to change sign is outside the
region we are modeling. The revised rotation profile alters only our
outermost annulus, through one data point which is averaged with its
counterpart on the opposite side of the galaxy. Were we to include the
revision, the value of $v_{225}$ in the
last row of Table \ref{t:data} would change from $53.1 \pm 26.9 \kms$ to
approximately $84 \pm 21 \kms$. This is a change of just over $1\sigma$
in one datum (with the second-largest error bar) out of 20 in the
velocity field. We have rerun models over part of the parameter space
with this change included and find only small differences; we will point
out below where the effect is noticeable.

\subsection{Results\label{s:results}}

Models are computed over essentially the same grid of dynamical parameters
used previously (\markcite{Sta94b}S94b, \markcite{Sta94c}S94c). At each of
the 9 pairs of $(k,l)$ values there are 8 choices of the ``contrast function''
(giving the relative streaming amplitude in short-axis and long-axis tubes),
4 choices of the function $v^\ast(t)$ (giving the angular dependence of the
mean velocity away from the symmetry planes), and two treatments of 
radial mean motions, with streamlines assumed to lie on either spherical
or ellipsoidal shells. The roles of the contrast and $v^\ast(t)$ functions
are summarized in Appendix B. We do not calculate models in which
the mass and luminosity distributions have different triaxialities since
our earlier tests showed that the triaxiality of the latter has little
bearing on that of the former. When we include the alignment and isophotal
twist constraints we implicitly assume the two triaxialities are equal; this
is justifiable {\it a posteriori\/} since nearly axisymmetric
potentials turn out strongly preferred. By Monte Carlo sampling over
$(T,c,\Omega)$,\footnote{From here on we omit the $L$ subscript on
$c$, but it is still the axis ratio of the luminosity distribution.}
the likelihood $L(T,c,\Omega,{\bf d})$ for each annulus is computed
for each set of dynamical parameters ${\bf d}$, and the individual likelihoods
are combined by the methods described in Section \ref{s:shapemethod}.

\subsubsection{The ``Maximal Ignorance'' Estimate}

\begin{figure}[t]
{\hfill\epsfxsize=4.0in\epsfbox{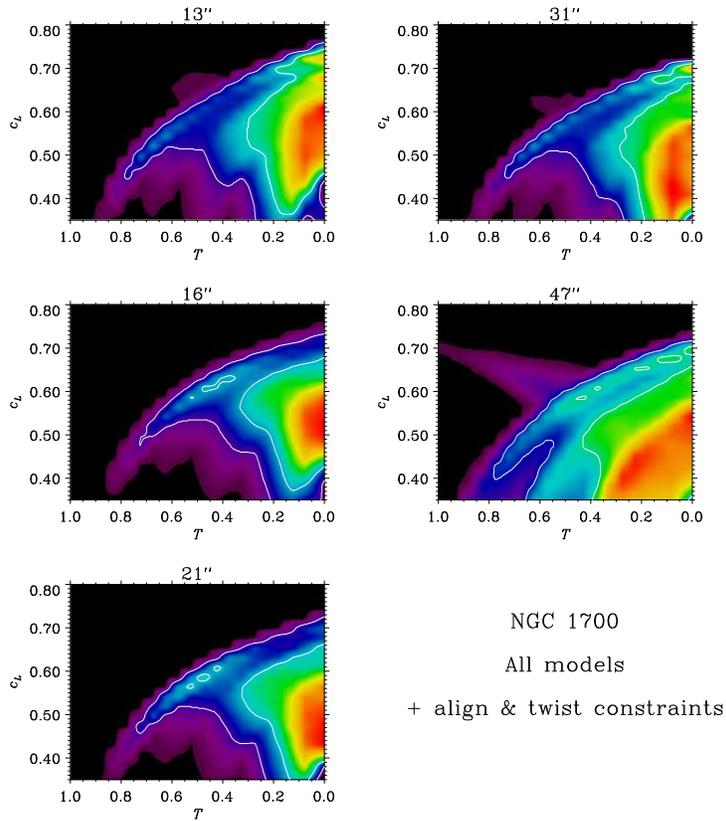}\hfill}
\caption{\footnotesize
``Maximal ignorance'' estimate of the intrinsic shape of NGC 1700 in five
radial zones. Each panel shows the marginal posterior density for the
triaxiality $T_i$ and short-to-long axis ratio $c_i$ of the $i$th annulus,
assuming that there is no intrinsic twist of the principal axes. Oblate
spheroids are at the right margin and prolate spheroids are at the left. All
velocity field models are used in this estimate, placing as few restrictions as
possible on the dynamics. White contours enclose the $68\%$ and $95\%$
higest posterior density (HPD) regions. Note the strong similarity of the
distributions for the inner four zones, the gradual tendency for greater
flattening with increasing radius, and the marked change in the outermost
annulus.
\label {f:allcolor}}\end{figure}

The marginal posterior densities $P_{12345}(T_i,c_i)$, including the
assumption of intrinsic alignment and the isophotal twist constraint
({\em cf.\/} equation \ref{e:generalbayes}), are shown in Figure
\ref{f:allcolor} for all models. This should be
considered the ``maximal ignorance'' estimate of intrinsic shape, since as
few restrictions as possible are placed on the dynamics.
In each panel, oblate spheroids
are at the right margin and prolate spheroids at the left. We rather
conservatively allow values of $c$ down to $0.35$, but since no models
with $c>0.8$ are consistent, at even the $4\sigma$ level, with the observed
$\epsilon >0.26$, the upper part of the parameter space is truncated. White
contours enclosing $68\%$ and $95\%$ of the total probability are the
Bayesian equivalent of $1\sigma$ and $2\sigma$ error bars.

The features to note in Figure \ref{f:allcolor} are: (1) the strong
similarity of the distributions for the inner 4 annuli, with prominent
peaks at $T=0$; (2) the overall trend toward smaller $c$ with increasing
radius, consistent with the gradual radial increase in $\epsilon$;
and (3) the significantly broader distribution in the outermost
annulus, indicating an increase in $T$. The diagonal ridges in the distribution
and the scalloped contours toward smaller $c$ are numerical artifacts
arising from undersampling of the sharply peaked densities
$P_i(T_i,c_i,\Omega_i)$.  Changing the outermost velocity on PA 225 (Sec.
\ref{s:forbidden}) causes a minuscule shift of the distributions
for the inner four annuli toward smaller $T$ and larger $c$, and in the
fifth annuls flattens the secondary maximum near $T=0.2$, $c=0.45$ and
truncates the weak appendage extending toward $T=1$, $c=0.7$.

\begin{figure}[t]
{\hfill\epsfxsize=4.0in\epsfbox{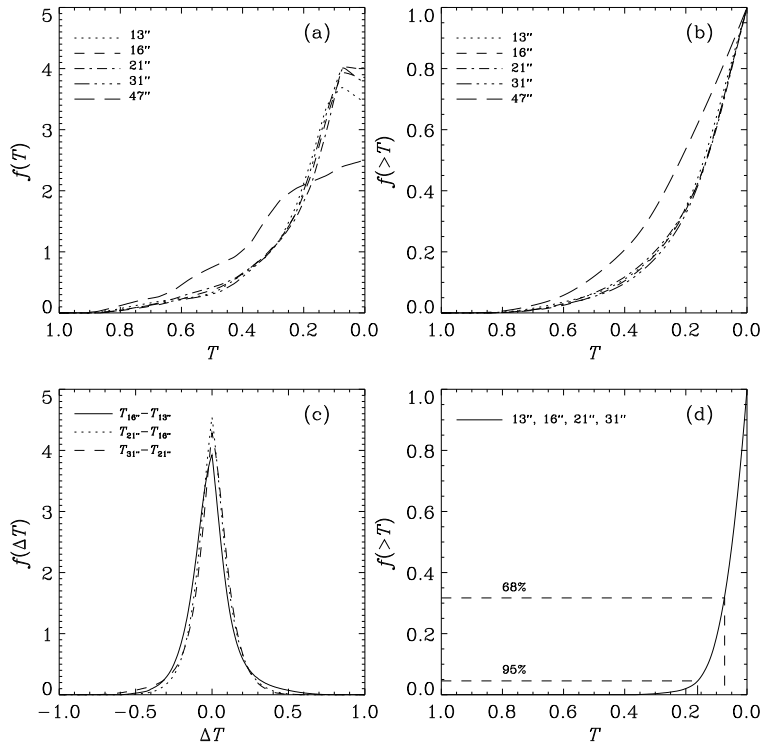}\hfill}
\caption{\footnotesize
($a$) Differential and ($b$) cumulative triaxiality distributions obtained
by integrating the posterior densities in Figure \protect{\ref{f:allcolor}}
over $c$. The nearly identical distributions in the inner 4 annuli imply
constant $T$ over this region, as do ($c$) the differential probabilities
for the change in triaxiality between adjacent annuli.
Including the assumption of constant $T$ in the analysis gives the
cumulative distribution for the single value of $T$ shown in ($d$).
Dashed lines indicate $1\sigma$ and $2\sigma$ upper limits.
\label {f:alltdist}}
\end{figure}

Figure \ref{f:allcolor} suggests that NGC 1700 is likely to be close to
axisymmetric, at least in the inner $35\arcsec$ or so. Considering its
central cuspiness, it is important to derive a quantitative constraint on
the triaxiality $T$. By integrating the distributions in Figure
\ref{f:allcolor} over $c$, we obtain the marginal densities
for $T$ shown in the first two panels of
Figure \ref{f:alltdist}. Figure \ref{f:alltdist}$a$ shows
the differential distributions; the corresponding
cumulative distributions are plotted in Figure \ref{f:alltdist}$b$.
The distributions for the inner 4 annuli are so similar that one is naturally
led to suspect that $T$ may be constant over this range of radii. Indeed,
plotting the integrated distributions tends somewhat to mask a strong
preference in the {\em joint\/} distribution for constant $T$. This is
demonstrated in Figure \ref{f:alltdist}$c$, in which we plot the differential
probabilities for the {\em change\/} in triaxiality between annuli 1 and 2,
2 and 3, and 3 and 4. All are sharply peaked at $\Delta T = 0$.
The strong similarity of these distributions and the small scatter in mean
values are consistent with $\Delta T$ being drawn from a
parent distribution that is a delta function at $\Delta T = 0$, indicating
constant $T$ in the inner 4 annuli.

If $T$ is {\em assumed\/} to be the same in annuli 1 -- 4, then this
assumption can be included as ``prior information'' (essentially 
by inserting factors of the form
$\delta(T_i-T_{i+1})\delta(T_{i+1}-T_{i+2})\cdots$ in equation
[\ref{e:generalkernel}]). The resulting cumulative distribution for the
single value of $T$ is shown in figure \ref{f:alltdist}$d$. Upper limits
on $T$ at the $1\sigma$ and $2\sigma$ confidence levels are drawn as
dotted lines. We obtain
\beq
\label{e:tlimit}
T < 0.16 \quad {\rm for}\quad 2.2h^{-1}\kpc < r < 6.7h^{-1}\kpc
\quad (95\% \rm{\ confidence}).
\eeq
A more intuitive but less precise statement is that the middle-to-long
axis ratio $b/a$ is greater than $0.9$ at $>95\%$ confidence.

\subsubsection{Statistical Effects of Orientation}

Because $P(T,c)$ is a marginal density integrated over orientation
$\Omega$, its value at a given $(T,c)$ is proportional to the
amount of solid angle over which a model of that shape will be seen to
reproduce the observations (averaged over the other parameters).
Consequently, there is a complementarity between constraints on shape and
orientation. Shapes able to fit the data
in many orientations are statistically favored while those requiring special
orientations are not; orientations that are required
by many shapes earn high marginal probabilities
while those useful to only a few score low. This leads to the
somewhat counterintuitive result that constraining the models
to be at the same orientation has the effect of narrowing
the marginal distribution in orientation while {\em broadening\/} that
in intrinsic shape. This effect can be rigorously demonstrated for analogous
one-dimensional problems. For NGC 1700's highly symmetric velocity field,
triaxial models requiring lines of sight in or near the $xz$ plane
to suppress apparent minor axis rotation are made relatively more likely by
the alignment constraint than axisymmetric models which can be more freely
oriented. The resulting marginal densities $P(\Omega)$ thus show a strong
preference for such orientations.

One should remember, however, that
this estimate is made in isolation, nearly independent of
knowledge of the rest of the elliptical galaxy population. That the
data may imply a special orientation for a single object is hardly alarming.
It is only in conjunction with other objects that the statistics are
affected by the requirement that the observed sample must be consistent
with a random drawing from an isotropic parent distribution. Since most
observed ellipticals are known not to have strongly triaxial-like kinematics,
this added constraint will push down the peak in the orientation distribution,
eliminating triaxial models that require that special orientation and
narrowing the marginal distribution for $T$. Methods for including
such constraints have been developed (Bak \& Statler \markcite{BaS97}1997)
and will be discussed fully and applied in a forthcoming paper
(Bak \& Statler \markcite{BaS98}1998, in preparation). For now
it suffices to acknowledge that including such constraints
in the modeling of NGC 1700 would tighten the limits on $T$, so that the
limit in equation (\ref{e:tlimit}) can be regarded as conservative.

\subsubsection{The Dynamical Prior\label{s:dynprior}}

The maximal ignorance estimate intentionally places as few restrictions as
possible on the orbit populations, because we have yet to understand how
Nature chooses to populate orbits. Such knowledge may be
forthcoming, however, either from observations or numerical simulations.
Prior restrictions on the dynamics will alter the shape estimate.

\begin{figure}[t]
{\hfill\epsfxsize=4.0in\epsfbox{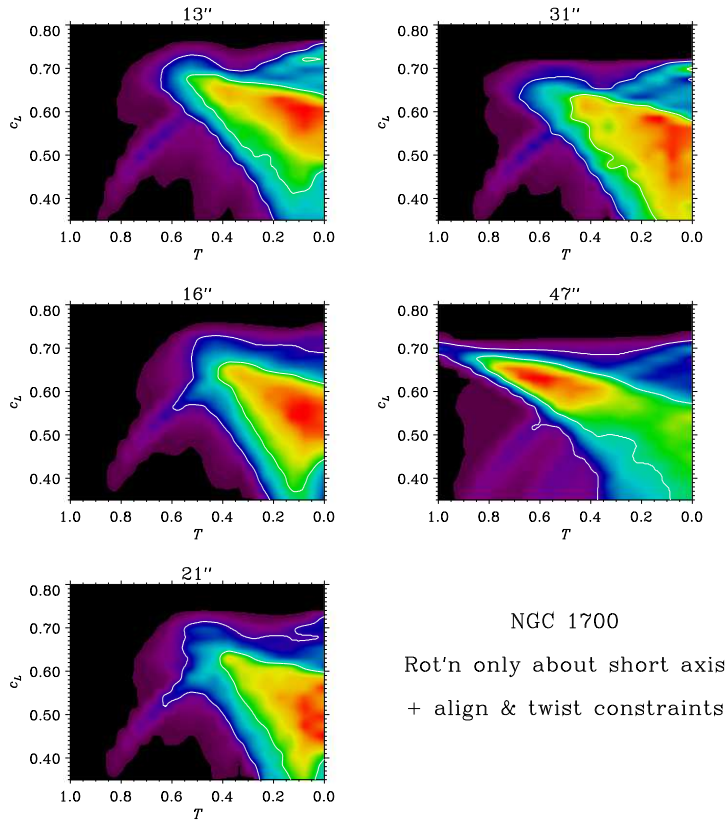}\hfill}
\caption{\footnotesize
Intrinsic shape estimate as in figure \protect{\ref{f:allcolor}}, but
excluding all models with net streaming in long-axis tubes. The galaxy
is inferred more likely to be triaxial, since triaxial systems forced to rotate
only about their short axes are more likely to be seen in orientations
where the velocity gradient along the apparent minor axis is zero.
\label {f:zonlycolor}}
\end{figure}

\begin{figure}[t]
{\hfill\epsfxsize=4.0in\epsfbox{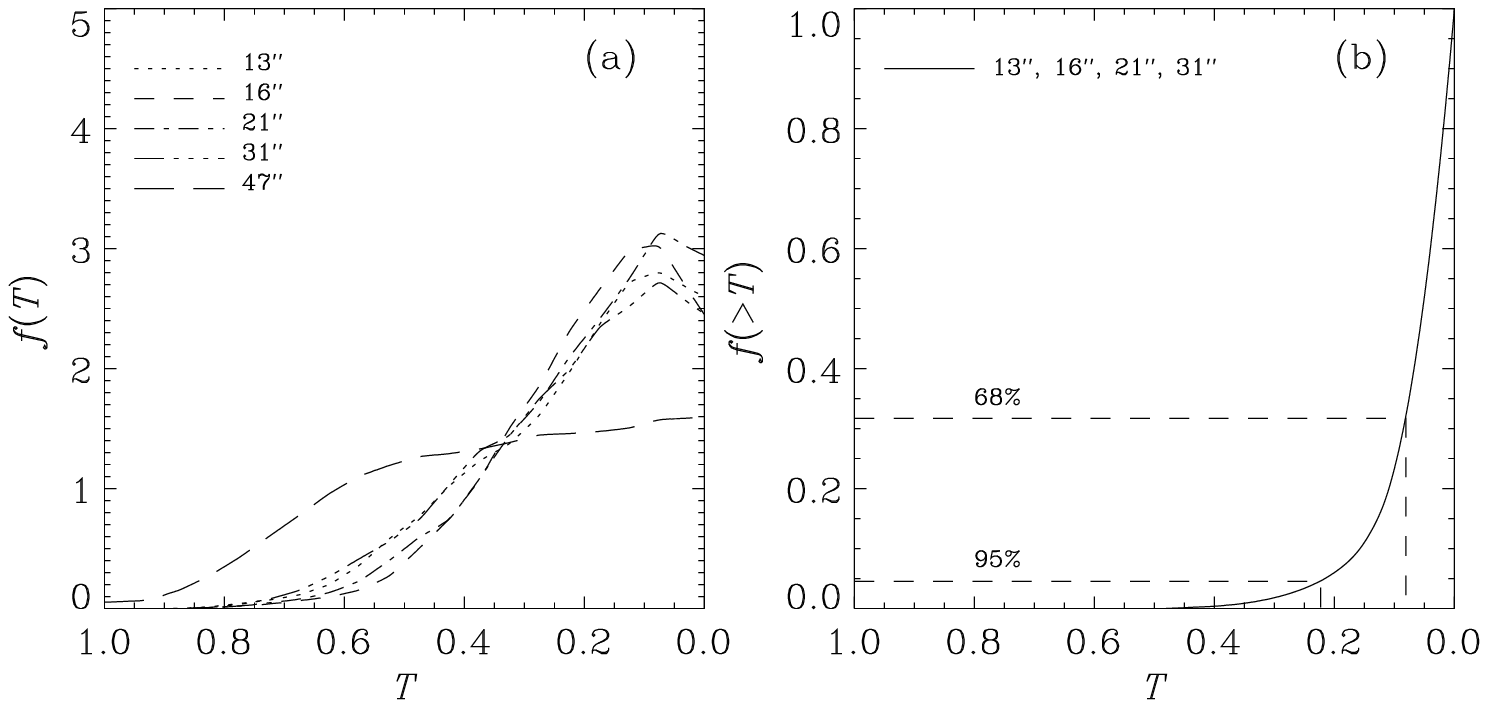}\hfill}
\caption{\footnotesize
($a$) Differential triaxiality distributions as in Figure
\protect{\ref{f:alltdist}}$a$, but for the case with net rotation in long-axis
tubes omitted. ($b$) Cumulative distribution for $T$ assuming the value
is constant in the inner 4 annuli. Dashed lines indicate $1\sigma$ and
$2\sigma$ upper limits.
\label {f:zonlytdist}}
\end{figure}

The infrequent occurrence of minor axis rotation in ellipticals seems to
indicate that the ``intrinsic misalignment'' between the angular momentum
vector and the short axis tends to be small (Franx, Illingworth, \& de Zeeuw
\markcite{FIZ91}1991). It may be the case that net streaming motions are
present only in the short-axis tube orbits.
If all models with long-axis tube streaming are excluded from
the fits, the result is the set of posterior densities shown in Figure
\ref{f:zonlycolor}. The galaxy comes out more triaxial, since, with
long-axis tube streaming disallowed, triaxial systems are more likely to
be seen in orientations where the apparent minor axis rotation is zero.
In Figure \ref{f:zonlytdist}$a$ we show the marginal densities for $T$ alone.
Those for the inner four annuli are not quite the same, but the
differences, compared with Figure \ref{f:alltdist}$a$, are consistent
with the larger amount of numerical noise in
the restricted set of models. Adding the prior constraint that $T$ is
constant over this range of radii gives the cumulative distribution in Figure
\ref{f:zonlytdist}$b$, and a 95\%-confidence upper limit of $T<0.22$. Note
that this is a change of only $0.06$ from the maximal ignorance limit.

\begin{figure}[t]
{\hfill\epsfxsize=4.0in\epsfbox{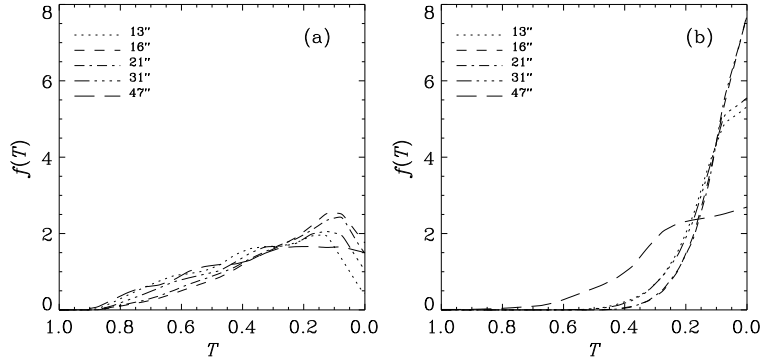}\hfill}
\caption{\footnotesize
Differential triaxiality distributions resulting from different choices of
the form of the $v^\ast(t)$ function (Appendix B). ($a$) Types 1 and 4; ($b$)
types 2 and 3. (See Section \protect{\ref{s:dynprior}}.)
\label {f:vstartdist}}
\end{figure}

By far the most important dynamical parameter is not
the relative loading of long- and short-axis tubes, but the form of
the $v^\ast(t)$ function, which gives the latitude dependence
of the mean streaming velocity across the $xz$ plane. Restricting the models to
$v^\ast(t)$ types 1 and 4 or to types 2 and 3
gives the marginal $T$ probabilities shown in figure \ref{f:vstartdist}.
(No restriction is placed on long-axis tube streaming.) The former
case shows a marked preference for triaxial shapes over axisymmetric
ones, the latter just the opposite.

The reason for this difference is the way in which the models fit the
position-angle dependence of the observed velocity field. SSC find that
velocities on the sampled PAs are consistent with an almost
exactly sinusoidal dependence at a given radius. To
understand how this affects the models, we note first that the results
for types 1 and 4 separately are nearly the same, as are those for types
2 and 3; this similarity indicates that the model velocity fields are
dominated by short-axis tubes. In triaxial systems, the sequence of
short-axis tubes terminates in the $xz$ plane at a latitude $\cos^{-1}T^{1/2}$
up from the $xy$ plane. The streamlines diverge to either side of this
constriction (see fig. 2b of \markcite{Sta94a}S94a). The mean flow in
the vicinity of the short axis will be seriously diluted, either by this
divergence alone, or by additional geometrical factors if the line of sight is
not close to the $xz$ plane --- {\em unless\/} either the flow velocity through
the constriction is sufficiently large near the terminating latitude, or $T$
is sufficiently small that the terminating latitude is itself close to the $z$
axis. The type 2 and 3 models, where the flow velocity through the
constriction falls off smoothly to zero with increasing latitude, rotate
too slowly on the diagonal position angles relative to the major axis
in projection, unless they are so close to oblate that the 
short-axis tubes extend nearly all the way to the $z$ axis. Conversely,
the type 1 and 4 models, where the velocity is assumed constant right up
to the terminating latitude, rotate too fast on the diagonal PAs
unless the terminating latitude is rather low, meaning $T$ is
significantly above zero.

Knowing the actual rate of occurrence of triaxiality in ellipticals has
major consequences for the abundance of steep central cusps and/or massive
black holes, and is an important test of the idea that central
mass concentrations can destroy triaxiality. The above result implies that
triaxiality constraints could be considerably strengthened if we could
restrict, on physical grounds, the form of $v^\ast(t)$. It is, therefore,
extremely important to obtain the $v^\ast(t)$ function from simulated
ellipticals, \eg\ from $N$-body merger experiments (Dutta, Statler,
\& Weil \markcite{DSW98}1998). Furthermore, understanding how this function
is determined physically by {\em different\/} mechanisms could make it
possible, with a sufficiently large sample, to use present-day kinematics
to constrain galaxy formation physics.

\section{Mass-to-Light Ratio and Anisotropy}

Given the upper limit on $T$ derived in the previous section, we can safely
use axisymmetric models to estimate the mass-to-light ratio in NGC 1700. 
A wealth of techniques have been developed over the past several years to fit
axisymmetric models to observations, of which two suffice for our purposes.

\subsection{Moment Models}

\subsubsection{Method}

For axisymmetric systems in which the phase space distribution function $f$
depends only on energy and the $z$ component of angular momentum, the
solution of the Jeans equations for the low order moments of $f$ can
be broken down into one-dimensional integrals along lines parallel
to the symmetry axis and derivatives of these integrals (Satoh
\markcite{Sat80}1980; Binney, Davies,
\& Illingworth \markcite{BDI90}1990, hereafter BDI). Efficient methods for
constructing such models were developed by \markcite{BDI90}BDI
and refined by van der Marel, Binney, \& Davies \markcite{MBD90}(1990;
hereafter MBD) and Magorrian \& Binney \markcite{MaB94}(1994).
Our implementation essentially parallels that of \markcite{BDI90}BDI and
\markcite{MBD90}MBD, with minor modifications.

The $R$-band surface
photometry is first interpolated onto an elliptic grid, spaced uniformly
in position angle $\psi$ and logarithmically in elliptic
radius $m_p = R (\cos^2\psi + \sin^2\psi/c_p^2)^{1/2}$; here $R$ is projected
radius on the sky and we take $c_p=0.7$. Since we are assuming
axisymmetry, PA twists are ignored. The photometry is extrapolated to
small and large radii using separate $r^{1/4}$-law fits to the inner and
outer parts of the radial profile.

The deprojection for a given
inclination $i$ starts with a parametric least-squares fit of a luminosity
density of the form
\beq
\label{e:initmodel}
\rho_L(m) = {\rho_0 a^3 \over m^2 (m^2 + a^2)^{1/2}}.
\eeq
In equation (\ref{e:initmodel}), $a$ is a scale length and $m$ is the
spheroidal radius $r (\cos^2\theta + \sin^2\theta/c^2)^{1/2}$, where
$c = [(c_p^2 - \cos^2 i)/\sin i]^{1/2}$ is the intrinsic axis ratio that
projects to the apparent ratio $c_p$ at inclination $i$.
This model projects to a surface brightness
\beq
I(m_p) = {I_0 a \over m_p} \tan^{-1}{a \over m_p},
\eeq
which is a better fit to the NGC 1700 data than the Hernquist and Jaffe
models used by \markcite{MBD90}MBD. The model is evaluated on a spheroidal
grid, and the fit is then improved by iterating
via Lucy's Method, giving a final luminosity model that is still axisymmetric
but not spheroidal. Three iterations are sufficient to bring
the photometric residuals down below $0.05$ magnitudes.

The mass density is found from
\beq
\rho(r,\theta) = \Upsilon\rho_L(r,\theta),
\eeq
where we can take the mass-to-light ratio $\Upsilon$ to be either constant
or a function of $m$. We obtain the gravitational potential
$\Phi$ by a spherical harmonic expansion, calculate the mean-square
velocities using BDI's equations (7) -- (8), and project them onto
the line of sight. We choose not to separate the ordered rotation from the
random tangential velocities since this introduces more free
parameters into the models. Instead, we add the observed dispersions and
mean velocities in quadrature and compare the resulting RMS with the
models.

\subsubsection{Results \label{s:momentresults}}

\begin{figure}[t]
{\hfill\epsfxsize=3.3in\epsfbox{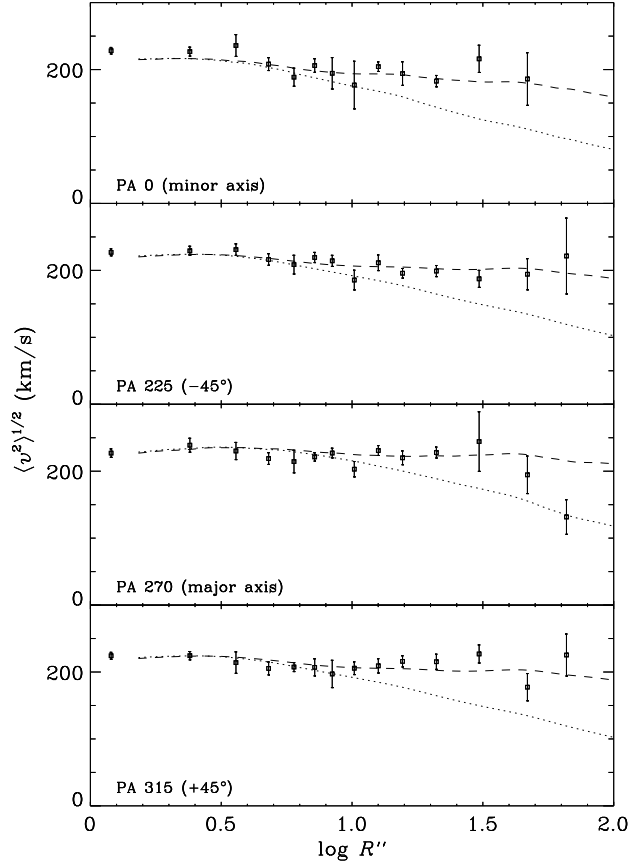}\hfill}
\caption{\footnotesize
The observed root-mean-square velocity profiles ({\em points with error
bars\/}) are compared with predictions of the the best 2-integral
constant-$M/L$ model ({\em dotted line\/}) and a 2-integral model with a
dark halo ({\em dashed line\/}). The constant-$M/L$ model fails miserably
while the halo model gives an acceptable fit. Both models are inclined
at $i=90\arcdeg$ (edge-on); models at lower inclinations are unable to match
the major to minor axis dispersion ratio.
\label {f:momentmodels}}
\end{figure}

The dotted lines in figure \ref{f:momentmodels} show the RMS velocity
profiles for the best constant-$M/L$ model. This model has $i=90\arcdeg$
(edge-on) and has been fitted by eye to the data in the inner $10\arcsec$.
The model clearly fails to reproduce the high RMS velocities observed beyond
$r_e$. Note also that for $R< 10\arcsec$, the velocity on the major axis
is slightly overestimated and that on the minor axis slightly
underestimated. This mismatch only gets worse for flatter models at lower
inclinations, since, to support the flattening, the tangential dispersion
must increase at the expense of the radial and vertical dispersions
(van der Marel \markcite{vdM91}1991). We conclude that 2-integral
models with constant $M/L$ cannot fit the data.

Two-integral models with variable $M/L$ fare better, though it is
difficult to constrain the details of the mass profile. To estimate the
magnitude of the required $M/L$ gradient, we adopt a simple law of the
form
\beq\label{e:upsilonofm}
\Upsilon(m) = \Upsilon_0 \left[1 + (m / r_h)^b\right]^{1/b},
\eeq
in which $\Upsilon$ is stratified approximately on density surfaces.
For large $b$, this describes an $M/L$ ratio that is constant at small
radii, turns up abruptly at $r_h$, and then increases linearly with radius.
Since the surface brightness is still falling roughly as $R^{-2}$ where the
photometry ends, $\Upsilon \propto r$ corresponds to what one expects from
a standard isothermal dark halo. Fixing the value of $b=8$ and minimizing
$\chi^2$ with respect to $\Upsilon_0$ and $r_h$ results in the model shown
as the dashed lines in figure \ref{f:momentmodels}; the fit is
statistically adequate, with a reduced $\chi^2$ of $1.16$ (49 degrees
of freedom). This model is inclined at $90\arcdeg$, since, as before,
lower inclinations are unable to match the major to minor axis dispersion
ratio. Making the halo spherical rather than spheroidal only exacerbates the
problem, since then the galaxy cannot rely on the flattening of the potential
to keep the luminosity density flat, and must become even more
tangentially anisotropic.

\begin{figure}[t]
{\hfill\epsfxsize=3.0in\epsfbox{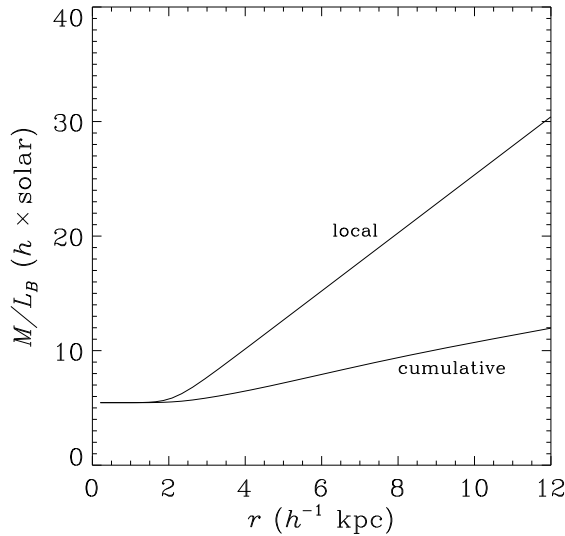}\hfill}
\caption{\footnotesize
Mass-to-light ratios, in terms of local density ratios and enclosed quantities,
plotted as functions of radius for the best 2-integral halo model shown in
Figure \protect{\ref{f:momentmodels}}.
\label {f:moverl}}
\end{figure}

The best fit local and cumulative $M/L$ profiles are shown in figure
\ref{f:moverl}. We have converted to blue luminosity using a mean $B-R$ of
$1.7$ \markcite{FIH89}(FIH). The central value of $5.5 h$ agrees very well
with Bender, Burstein, \& Faber's \markcite{BBF92}(1992) result
of $2.8$ for $h=0.5$, and the factor of $\sim 2$ cumulative increase out to
 a scale of $\sim 10 h^{-1} \kpc$ is consistent with standard halo models.
Thus 2-integral models with dark matter appear able to fit the data.

This result, however, is less than entirely satisfactory for at least three
reasons. First, the velocity field fitting of Section \ref{s:shape} implies
flatter shapes and lower inclinations than obtained here. Second, the
mere existence of a solution to the Jeans equations does not guarantee the
existence of a non-negative distribution function $f$. And third, since
mass can be traded off with anisotropy, 3-integral models are essential
to determine whether an $M/L$ gradient is actually {\em required\/} by the
data.

\subsection{Quadratic Programming Models}

\subsubsection{Method}

Three-integral models constructed by quadratic programming (QP) score over
moment models in that, since one solves explicitly for a positive definite
phase space distribution function, there is no risk that an apparently
successful model may require an unphysical negative density.

The numerical
approach used here is essentially the one described by Dejonghe \etal\
\markcite{Dej96}(1996). In brief, the procedure is as follows: first, for
an assumed inclination $i$, the surface photometry is deprojected as
described above using Lucy's method.
The total mass distribution is assumed to have the same {\em shape\/} as
the luminosity density, though not necessarily the same radial profile. This is
accomplished by setting, for each term in the spherical harmonic expansion
of the density,
\beq\label{e:qpupsilonofm}
\rho_{lm}(r) = \rho_{L,lm}(r)\Upsilon_0(1+ B r^p);
\eeq
the case of constant $M/L$ then corresponds to $B=0$.  The potential
is computed from $\rho$ by harmonic expansion. A St\"ackel fit is made
to the potential, using the de Zeeuw \& Lynden-Bell
\markcite{ZLB85}(1985) method as implemented by Dejonghe \& de Zeeuw
\markcite{DdZ88}(1988). This fit is made only to define an approximate
third integral $I_3$; wherever the potential is needed during the
modeling, the original potential is used. The distribution function
$f$ is built up as a sum of components of Fricke type:
\beq
f(E,L_z,I_3) = \sum_{p,q,n} A^\pm_{pqn} F^\pm_{pqn},
\eeq
where $n$ is an integer but $p$ and $q$ may be real, and
\beq
F^\pm_{pqn} = \left\{ \begin{array}{ll}
E^p (EL_z^2/2)^q (EI_3)^n, & \pm L_z \geq 0, \\
0 & \pm L_z < 0.
\end{array} \right.
\eeq
For components of this type, the velocity moments can be expressed in
terms of algebraic functions of the coordinates and the potential.
Models are constructed from a library of components with various combinations
of $p$, $q$ and $n$, generated by the direct product of 
$p=2\ldots10$, $q=0\ldots5$ and $n=0\ldots4$. The QP
algorithm is used to compute the coefficients $A^\pm_{pqn}$ that
optimize the fit to the photometric data and the mean velocity and
dispersion profiles. This is done by minimizing a $\chi^2$-like variable
of the form
\beq
\chi^2 = \sum_i w_i\left({\mu_{{\rm obs},i}-\mu_{{\rm mod},i}\over
\mu_{{\rm mod},i}}\right)^2
\eeq
where the $\mu$'s represent the photometric and kinematic data collectively and
the sum is over all data points. While complete two-dimensional surface
photometry is available, we choose for convenience to fit the photometry only
at the points where there are kinematic measurements. This allows some of the
models to come out a bit peanut-shaped, but the effect on the inferred mass
profile is minimal. The weights $w_i$ are in principle arbitrary, but are here
assigned the value 1.

\subsubsection{Results}

\begin{figure}[t]
{\hfill\epsfxsize=3.3in\epsfbox{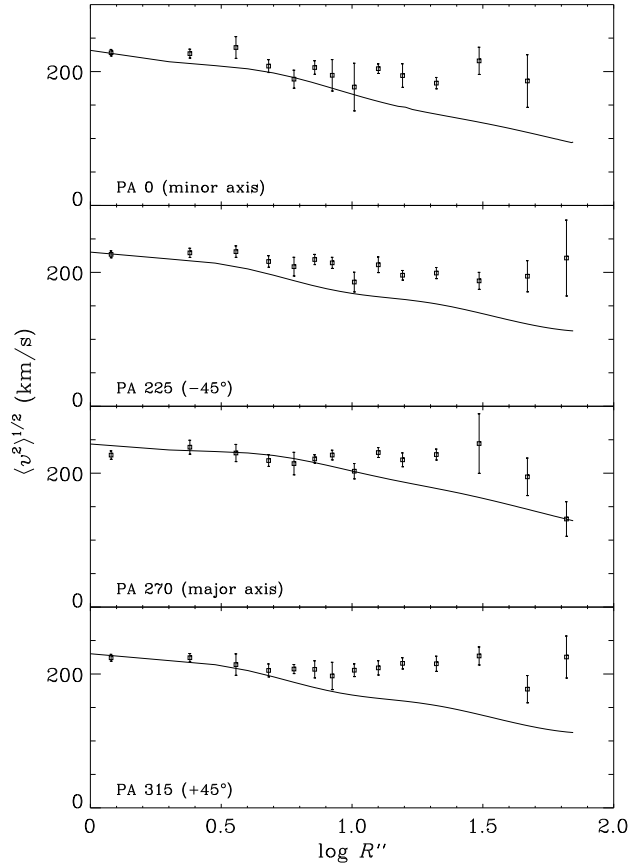}\hfill}
\caption{\footnotesize
The best three-integral constant-$M/L$ model compared with the observed
root-mean-square velocity profiles, as in figure \protect{\ref{f:momentmodels}}.
The added freedom in the distribution is insufficient to reproduce the
shallow profile at large radii.
\label{f:constmlmodel}}
\end{figure}

The best constant-$M/L$ 3-integral model, at an inclination of
$90\arcdeg$, is shown in figure \ref{f:constmlmodel}.
The global value of $M/L_B = 2.8 h$ has been chosen to optimize the overall
fit, \ie\ without assigning larger weight to any particular region, though
the higher $S/N$ at smaller radii makes the model appear pinned to the
central dispersion. The shallow slope of the RMS velocity profile cannot be
reproduced; thus, allowing for anisotropy does not
eliminate the need for an $M/L$ gradient. In fact, the fit is even slightly
worse than for the 2-integral moment model of \S\ \ref{s:momentresults}.
We ascribe this to the positivity requirement on $f$ in the QP models, which is
not applied in the moment models. If the positivity constraint is lifted,
the same QP components can produce a fit with a $\chi^2$ a factor of 3 lower.
We take this as an indication that the constant-$M/L$ moment
model actually does not correpsond to a nonnegative distribution function. 

\begin{figure}[t]
{\hfill\epsfxsize=3.3in\epsfbox{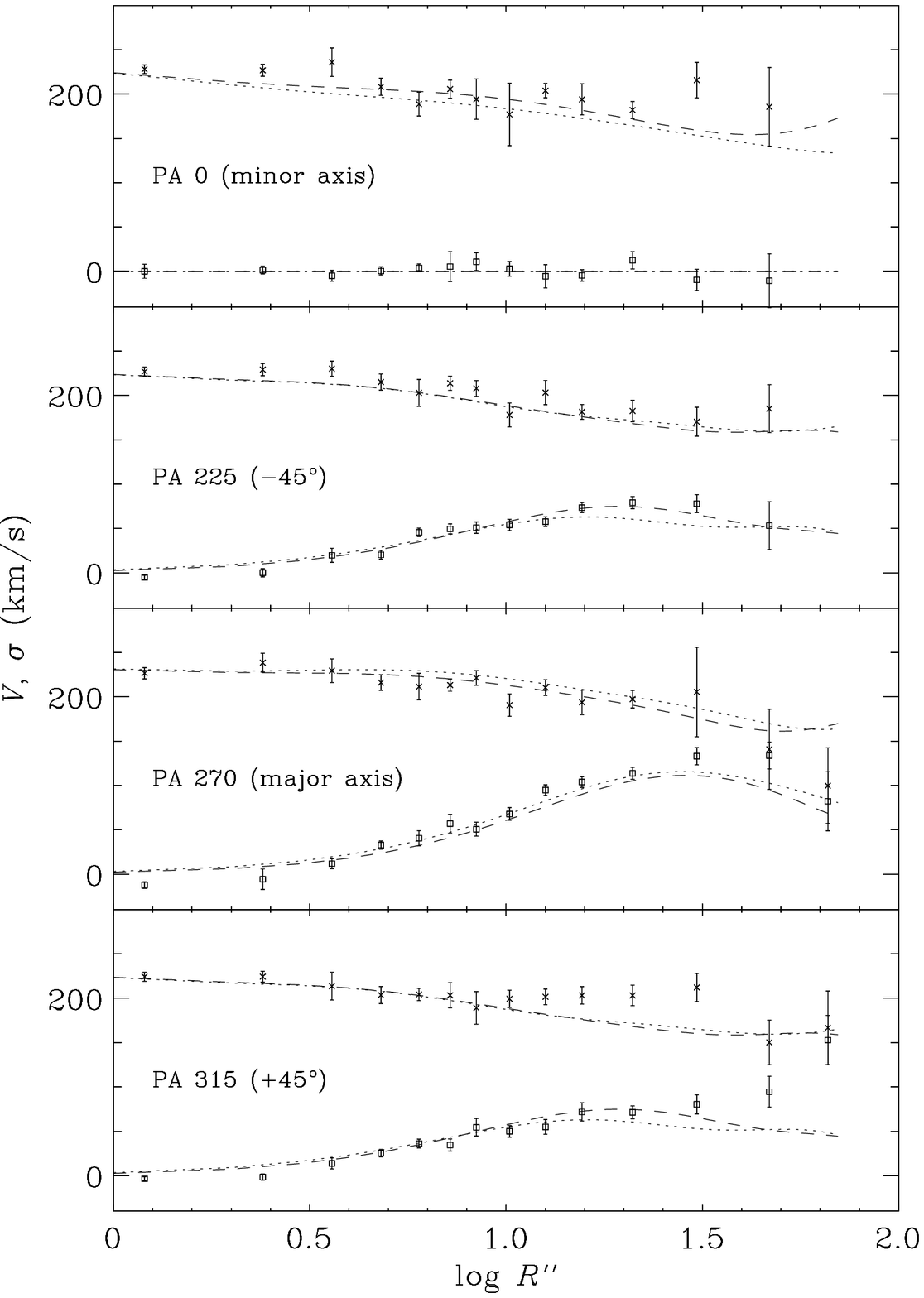}\hfill}
\caption{\footnotesize Three-integral, variable-$M/L$ models that adequately
fit the kinematic data, compared with the observed mean velocities
({\em squares}) and dispersions ({\em crosses}). Models have inclinations
of $90\arcdeg$ ({\em dashed lines\/}) and $60\arcdeg$ ({\em dotted lines\/}).
The $60\arcdeg$ model has been scaled up by 3\% from the formal best fit to
better reproduce the central dispersion.
\label{f:qpmodels}}
\end{figure}

Allowing for an $M/L$ gradient improves the fit substantially. The dashed
lines in figure \ref{f:qpmodels} show a model with $i=90\arcdeg$, $p=1$,
$B=1.27 h^{-1}\kpc^{-1}$, and $\Upsilon_0=5.1 h$ ($B$ band, in solar units),
compared with
the observed rotation (squares) and dispersion (crosses) profiles. The fit
to the kinematics alone, ignoring the photometry, is passable, with
$\chi^2/\nu = 2.03$ for $\nu=108$ degrees of freedom. The discrepancy
at large radii on PA 315 is not especially worrisome since this is where
isophotal and kinematic twisting sets in, and axisymmetric models should
not be expected to reproduce the detailed velocity field.

\begin{figure}[t]
{\hfill\epsfxsize=3.0in\epsfbox{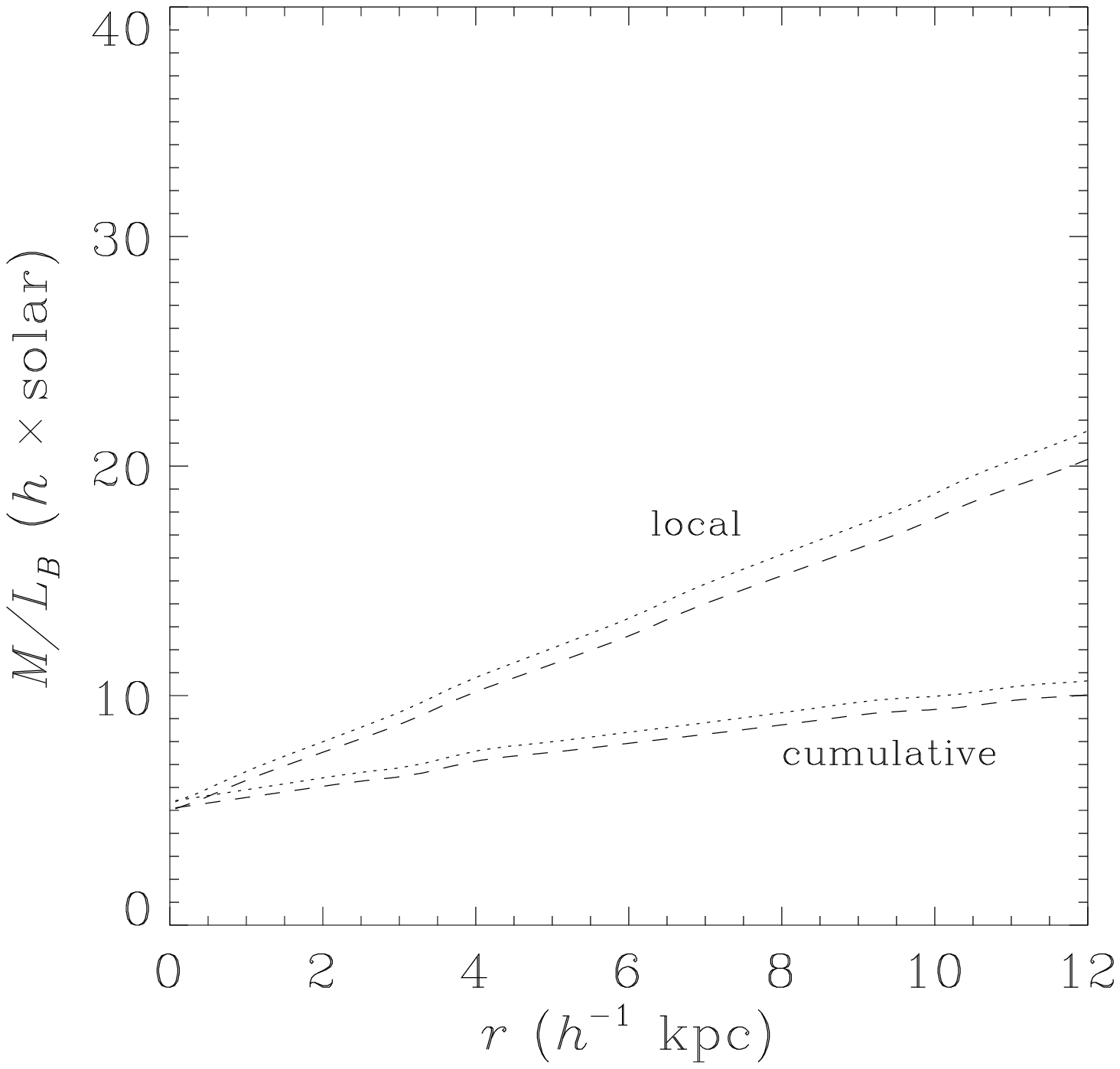}\hfill}
\caption{\footnotesize Local and cumulative mass-to-light ratios, as in
Figure \protect{\ref{f:moverl}}, for the $i = 90\arcdeg$ ({\em dashed lines\/})
model and the $i = 60\arcdeg$ model ({\em dotted lines\/}) shown
in Figure \protect{\ref{f:qpmodels}}.
\label{f:qpmoverl}}
\end{figure}

The local and cumulative $M/L$ profiles for this model are shown as the dashed
lines in figure \ref{f:qpmoverl}. The central values agree very well with
those shown in Figure \ref{f:moverl} for the 2-integral model, and appear
to be insensitive to assumptions about the functional form
(\ie\ equation [\ref{e:upsilonofm}] {\em vs.\/} equation
[\ref{e:qpupsilonofm}].) At large $r$ the local $M/L$ gradient is roughly
a factor $1.5$ smaller than that obtained from the 2-integral model. This
translates to a $\sim 20\%$ difference in the cumulative
$M/L$ gradient because of the steepness of the mass profile.

\begin{figure}[t]
{\hfill\epsfxsize=4.0in\epsfbox{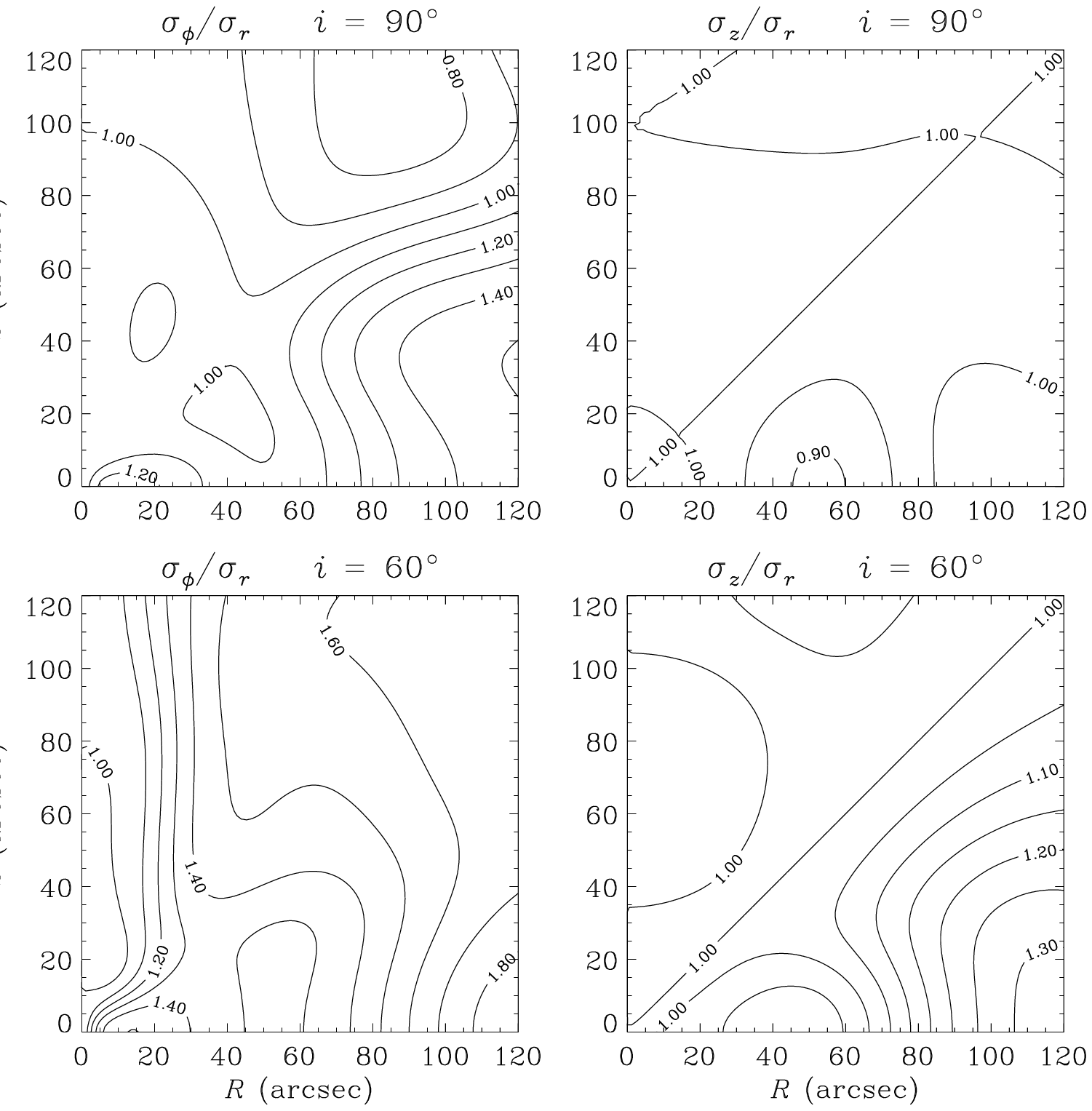}\hfill}
\caption{\footnotesize Ratios of the internal (\ie\ not projected) velocity
dispersions for the ({\em top}) $i = 90\arcdeg$ and ({\em bottom})
$i = 60\arcdeg$ models of Figure \protect{\ref{f:qpmodels}}. The ratios of
the ({\em left}) azimuthal and ({\em right}) vertical components of 
dispersion to the radial component are contoured in the meridional plane.
Kinematic data end at around $70\arcsec$.
\label{f:qpdispersions}}
\end{figure}

Obtaining a strict lower limit on the $M/L$ gradient, while desirable,
is not especially practical. A shallower gradient can be obtained at the
expense of a more pathological distribution function at low binding energy,
corresponding to distortions of the LOSVD at radii for which there are no
data. As there is no objective criterion for accepting or rejecting such
models, we have rather subjectively chosen the parameters $B$ and $p$ so
as not to overly tax the QP algorithm; thus the distribution function
is not on the verge of going negative,
and the anisotropy is not extreme. The latter property is illustrated in
the top two panels of figure \ref{f:qpdispersions}, which show the ratios
of the unprojected dispersions $\sigma_R$, $\sigma_\phi$, and $\sigma_z$
in the meridional plane. The velocity ellipsoids are generally elongated
in the tangential direction, and $\sigma_z / \sigma_R$ is within $10\%$
of unity nearly everywhere. The model becomes more tangential at larger
$R$; but remember that the kinematic data end near $70\arcsec$, so the
velocity distribution beyond this point is not well constrained.

Though we have not checked explicitly, it seems virtually certain that for
any inclination $i>50\arcdeg$ (for which $c>0.35$) it is possible
to find many values of $B$ and $p$ that produce an acceptable fit.
For the shape fits of Section \ref{s:shape}, the posterior densities in
figures \ref{f:allcolor} and \ref{f:zonlycolor} give a most-probable
intrinsic flattening $c$ between $0.55$ and $0.65$, which would correspond,
for an axisymmetric system, to $i \approx 60\arcdeg$. A QP
model at this inclination, using the same $B$ and $p$ as the $i=90\arcdeg$
model above, is shown as the dotted lines in figure \ref{f:qpmodels}.
The formal best fit has been scaled up by $3\%$ in velocity, or $6\%$ in
mass, to improve the fit to the central dispersion. The local and
cumulative $M/L$ profiles are plotted as the dotted lines in figure
\ref{f:qpmoverl}, and the ratios of the principal dispersions are shown
in the bottom two panels in figure \ref{f:qpdispersions}. This rather
flat model is tangential at all radii, and at large $R$ (past where the
data end) becomes radially quite cold. Such a configuration is not at
all impossible, especially if, as \markcite{SSC}SSC suggest, the main body
of the galaxy is surrounded by a differentially precessing disk acquired
through a merger or other accretion event.

\section{Discussion\label{s:discussion}}

By modeling its velocity field and surface photometry, we have found
that NGC 1700 is nearly axisymmetric and oblate. We can say with better
than $95\%$ confidence that if it were viewed down its short axis, it would
appear as at most an E1 at radii $\lesssim 36\arcsec$ ($6.7 h^{-1}\kpc$).
Beyond this point, if the density surfaces remain aligned, then photometric
and kinematic twists indicate that it becomes significantly flatter and more
triaxial.

The dominant source of systematic error in the determination of intrinsic
shape is our lack of prior knowledge about the internal dynamical structure,
specifically, how we should expect the mean streaming velocity to vary
away from the symmetry planes. Axisymmetry is strongly favored if the
velocity declines away from the $xy$ plane; triaxiality is preferred if
it doesn't. It is therefore important to understand how the latitude
dependence of the mean streaming may be determined by physics.
Preliminary results of an analysis of group
merger simulations (Dutta, Statler, \& Weil \markcite{DSW98}1998),
though limited by small-number statistics, suggest that the mean velocity
does decline away from the $xy$ plane in objects produced by this
mechanism. It remains to be seen whether different formation scenarios
would produce different behavior.

NGC 1700 falls into the family of steep-cusped ellipticals delineated by
WFPC photometry, standing out, in fact, as its most luminous member (so far).
In general, the steep-cusped galaxies show characteristics one would expect of
axisymmetric systems, such as disky isophotes and relatively rapid rotation.
The near axisymmetry we have found is consistent with Merritt's
\markcite{Mer97}(1997) result that orbital stochasticity prevents the
existence of self-consistent triaxial equilibria for the steep ($r^{-2}$)
cusped Jaffe profile unless $T\lesssim 0.4$, $T \gtrsim 0.9$, or $c \gtrsim
0.8$. It is tempting to interpret the change of shape beyond $40\arcsec$
as marking the limit out to which triaxiality has been eaten away by
stochastic scattering of box orbits. But it is far from clear that the cusp
in NGC 1700 is the cause of this structure. As discussed by \markcite{SSC}SSC,
there is reason to believe that the observed photometric and kinematic twists
arise from an outer accreted stellar disk or ring which is incompletely
phase-mixed and therefore intrinsically misaligned with the rest of the galaxy.
In this case the dynamical time at the location of the twist, times a factor
of order a few, indicates the time since the last significant merging event.
The dynamical age of $2$ to $4 h^{-1} \gyr$ \markcite{SSC}(SSC) is corroborated
by Whitmore \etal\ \markcite{Whi97}(1997), who have observed NGC 1700's
globular clusters and find colors and magnitudes consistent with a comparably
old cluster population.

We have also found, using axisymmetric 2-integral moment models and 3-integral
quadratic programming models, that NGC 1700 must have a radially increasing
mass to light ratio. No constant-$M/L$ model can fit the data.
The cumulative (enclosed) $M/L_B$ increases from a
central value of $\sim 5 h$ to $\sim 10 h$ (in solar units) at
$r = 12 h^{-1}\kpc$. The case for dark matter in NGC 1700 is strong; even
allowing for varying anisotropy, the best constant-$M/L$
model fits the RMS velocity profiles at no better than
$\chi^2/\nu=5.74$ (for $\nu=50$). In this sense the evidence of dark matter is
comparable to that in NGC 2434 (Rix \etal\ \markcite{Rix97}1997); in fact,
NGC 1700 is arguably a more compelling case since only spherical models have
been applied to NGC 2434.
Other systems suspected on stellar dynamical grounds of harboring
substantial dark halos, such as NGC 1399, 4374, 4472, 5813, 7626, 7796, and
IC 4296 (Saglia, Bertin, \& Stiavelli \markcite{SBS92}1992, Saglia
\etal\ \markcite{SKP93}1993, Bertin \etal\ \markcite{Ber94}1994)
are statistically less convincing. NGC 1700 appears to represent the strongest
stellar dynamical evidence to date for dark halos in ellipticals.

Bertin \etal\ \markcite{Ber94}(1994) point out that, of the ellipticals
listed above with stellar dynamical indications of dark matter, most turn out
to be known bright X-ray sources. Unfortunately, NGC 1700 seems never to
have been observed by an X-ray telescope, though ROSAT programs are pending.
NGC 1700 bears a more-than-passing resemblance to NGC 7626
(Saglia, Bertin, \& Stiavelli \markcite{SBS92}1992), having a very similar
luminosity, central dispersion, effective radius, and distance. However,
NGC 7626 is 2 orders of magnitude more powerful in the radio (Bender
\etal\ \markcite{Ben89}1989), and also resides in a rich cluster
while NGC 1700 has only one nearby companion and lies among small groups.
It will be very interesting to see whether these two apparently similar
ellipticals, in very different environments, turn out to be X-ray twins.

\acknowledgments

TSS acknowledges support of this work from NASA Astrophysical Theory Program
Grant NAG5-3050 and NSF CAREER grant AST-9703036.
We thank the director and staff of the Multiple
Mirror Telescope Observatory for their generous assistance and allocations
of time to this project, and Trevor Ponman and Joe Shields for timely
information.

\section*{Appendix A: Combining Shape Estimates for Multiple Annuli}

We begin by introducing the notation $P_{ijk\ldots}(T_i,c_i,\Omega_i)$ for
the marginal probability density for the shape and orientation of annulus
$i$, making use of the data from annuli $i,j,k,\ldots$. Suppose that we have
obtained $P_1(T_1,c_1,\Omega_1)$ and $P_2(T_2,c_2,\Omega_2)$ for annuli 1 and 2
individually (\cf\ equation [\ref{e:margfourd}]). (Here {\em both\/} $T$ and
$c$ refer to the light distribution; we omit the $L$ subscripts for
clarity.) We assume that the luminosity
distribution is intrinsically aligned ($\Omega_1=\Omega_2=\Omega$)
and require that the observed position angle difference
$\delta_{12} \pm \sigma_{\delta_{12}}$ be reproduced. Let
$L_{\delta_{12}}(T_1,T_2,\Omega)$ be the likelihood of the observed
twist given the pair of intrinsic shapes $(T_1,c_1)$ and $(T_2,c_2)$, and
orientation $\Omega$:
\beq
\label{e:ldelta}
L_{\delta_{12}}(T_1,T_2,\Omega) = \exp \left\{-
{[\delta_{12}-\delta_{12}^{\rm mod}(T_1,T_2,\Omega)]^2 \over 2
\sigma_{\delta_{12}}^2}
\right\}.
\eeq
The axis ratios $c_1$ and $c_2$ do not appear in either the likelihood
$L_{\delta_{12}}$ or the predicted twist $\delta_{12}^{\rm mod}$ because
in the quasi-local approximation the major axis position
angle in annulus $i$,
\beq
{\rm PA}_i = {1 \over 2} \tan^{-1} {T_i \cos \theta_E \sin 2 \phi_E
\over | T_i(\sin^2 \phi_E - \cos^2 \phi_E \cos^2 \theta_E) - \sin^2 \phi_E |}.
\eeq
is independent of $c_i$. The joint
probability that the two shapes fit the local photometry and
kinematics {\em and\/} the PA twist is given by
\beq
P(T_1,c_1,T_2,c_2) = \int d\Omega\, P_1(T_1,c_1,\Omega) P_2(T_2,c_2,\Omega)
L_{\delta_{12}}(T_1,T_2,\Omega),
\eeq
but this is a somewhat unwieldy quantity. Of more use is the
marginal probability for, say, annulus 1 alone:
\beq
P_{12}(T_1,c_1) = \int dT_2\, dc_2\, P(T_1,c_1,T_2,c_2)
= \int d \Omega\, P_1(T_1,c_1,\Omega) K_{12}(T_1,\Omega),
\eeq
where the ``twist kernel'' $K_{12}$ is given by
\beq
K_{12}(T_1,\Omega) = \int d T_2\, L_{\delta_{12}}(T_1,T_2,\Omega)
P_2(T_2,\Omega),
\eeq
and $P_2(T_2,\Omega)$ is the marginal distribution integrated over $c_2$, \ie
\beq
P_2(T_2,\Omega) \equiv \int dc_2\, P_2(T_2,c_2,\Omega).
\eeq
The posterior $P_{12}(T_1,c_1)$ is the shape estimate for annulus 1
considering all of the data in both annuli, but regardless of the actual
shape of annulus 2.

In our case the $P_i(T_i,c_i,\Omega_i)$ are obtained from the dynamical
models, so the $T_i$ are mass triaxialities. If one is unwilling to assume
a relation between the triaxialities of mass and luminosity, one can
still combine the annular fits, assuming only that the principal axes are
intrinsically aligned and foregoing the isophotal twist constraint, by
letting $\sigma_{\delta_{12}} \to
\infty$, which gives $L_{\delta_{12}}=1$ by equation (\ref{e:ldelta}).

The generalization to $N$ annuli uses the measured PA
gradients $\delta_{12}, \delta_{23},\ldots,\delta_{N-1,N}$, as follows:
\beq
\label{e:generalbayes}
P_{12 \ldots N}(T_i,c_i,\Omega) = P_i(T_i,c_i,\Omega)
K^{\rm int}_{i,i-1,\ldots,1}(T_i,\Omega)
K^{\rm ext}_{i,i+1,\ldots,N}(T_i,\Omega),
\eeq
where the interior and exterior kernels are given by
\begin{mathletters}
\label{e:generalkernel}
\begin{eqnarray}
K^{\rm int}_{i,i-1,\ldots,1}(T_i,\Omega) &=&
	\int d T_{i-1}\, P_{i-1}(T_{i-1},\Omega)
	L_{\delta_{i-1,i}}(T_{i-1},T_i,\Omega) \nonumber \\
& & \phantom{\cdots}\int d T_{i-2}\, P_{i-2}(T_{i-2},\Omega)
	L_{\delta_{i-2,i-1}}(T_{i-2},T_{i-1},\Omega) \cdots \\
& & \phantom{\cdots\cdots}\int d T_{1}\, P_1(T_1,\Omega)
	L_{\delta_{1,2}}(T_1,T_2,\Omega); \nonumber \\
K^{\rm ext}_{i,i+1,\ldots,N}(T_i,\Omega) &=&
	\int d T_{i+1}\, P_{i+1}(T_{i+1},\Omega)
	L_{\delta_{i,i+1}}(T_i,T_{i+1},\Omega) \nonumber \\
& & \phantom{\cdots}\int d T_{i+2}\, P_{i+2}(T_{i+2},\Omega)
	L_{\delta_{i+1,i+2}}(T_{i+1},T_{i+2},\Omega) \cdots \\
& & \phantom{\cdots\cdots} \int d T_{N}\, P_N(T_N,\Omega)
	L_{\delta_{N-1,N}}(T_{N-1},T_N,\Omega). \nonumber
\end{eqnarray}
\end{mathletters}
As before, the shape estimate for annulus $i$, using all of the data but
regardless of the shape at other radii, is the marginal distribution
$P_{12 \ldots N}(T_i, c_i) = \int d\Omega\, P_{12 \ldots N}(T_i, c_i, \Omega)$.
The construction of equations (\ref{e:generalbayes}) and
(\ref{e:generalkernel}) guarantees that the
marginal distribution in orientation,
$P_{12 \ldots N}(\Omega) = \int dT_i\, dc_i P_{12 \ldots N}(T_i, c_i, \Omega)$,
is independent of $i$, as it must be given the assumption of intrinsic
alignment.

\section*{Appendix B: The Contrast and $v^\ast(t)$ Functions}

In the models used in Section \ref{s:shape}, properties of the phase space
distribution function are subsumed into the ``similar flow'' ansatz, a
scalar constant $C$, and a function of one variable $v^\ast(t)$. The last
two describe the mean velocity across the $xz$ plane on one fiducial
spherical or ellipsoidal shell, which in turn determines the velocity
field over the whole shell once $T$ and the luminosity density are
specified.

The ``contrast'' $C$ is defined as the ratio of the $y$ component of the
mean velocity on the $x$ axis to that on the $z$ axis, on the fiducial shell.
One may think of this as the ratio of the maximum mean streaming velocities
produced by short-axis and long-axis tubes. Models lacking short-axis or
long-axis tube streaming have $C=0$ or $C=\infty$, respectively. Models
are run with values of $C$ both independent of intrinsic shape and given
by the four functions of $(T,c)$ shown in Figure 2 of \markcite{Sta94c}S94c.

The function $v^\ast(t)$ gives the angular dependence of the mean velocity
across the $xz$ plane on the fiducial shell. The variable $t$ is a
rescaled polar angle, defined so that $t=0$ corresponds
to the $x$ axis, $t=2$ to the $z$ axis, and $t=1$ to the locus
dividing the short-axis from the long-axis tubes. For spherical shells,
\beq
t = \left\{ \begin{array}{ll}
        2 - {\sin^2 \theta \over T}, & \theta < \sin^{-1}\sqrt{T}, \\ 
        {\cos^2 \theta \over 1-T}, & \theta > \sin^{-1}\sqrt{T};
\end{array} \right.
\eeq
for ellipsoidal shells the relation between $t$ and $\theta$ is more
complicated, and we refer the reader to Section 3.1 of \markcite{Sta94c}S94c.
By definition $v^\ast(0)=C$ and $v^\ast(2)=1$. In the intervals
$[0,1)$ and $(1,2]$, $v^\ast(t)$is assumed to be either piecewise-constant
or piecewise linear, in the latter case falling to zero at $t=1$. This
gives four possible forms for $v^\ast(t)$: Type 1, constant in both
intervals; Type 2, linear in both intervals; Type 3, linear in $[0,1)$ and
constant in $(1,2]$; and Type 4, constant in $[0,1)$ and linear in $(1,2]$.


\begin{references}
\reference{AnS98} Anderson, R. F. \& Statler, T. S. 1998, \apj, 496, 706
\reference{BaS97} Bak, J. \& Statler, T. S. 1997, in The Second Stromlo
	Symposium: The Nature of Elliptical Galaxies, ed. M. Arnaboldi, G.
	S. Da Costa, \& P. Saha (San Francisco: Astronomical Society of the
	Pacific)
\reference{BaS98} Bak, J. \& Statler, T. S. 1998, in preparation
\reference{BBF92} Bender, R., Burstein, D., \& Faber, S. M. 1992, \apj,
	399, 462
\reference{Ben89} Bender, R., Surma, P., D\"obereinder, S., M\"ollenhoff, C.,
	\& Madejsky, R. 1989, \aap, 217, 35
\reference{BDI90} Binney, J. J., Davies, R. L., \& Illingworth, G. D. 1990,
	\apj, 361, 78 (BDI)
\reference{Ben89} Bender, R., Surma, P., D\"obereiner, S., M\"ollenhoff,
	C., \& Madejsky, R. 1989 \aap, 217, 35
\reference{Ber94} Bertin, G., Bertola, F., Buson, L. M., Danziger, I. J.,
	Dejonghe, H., Sadler, E. M., Saglia, R. P., de Zeeuw, P. T., \&
	Zeilinger, W. W. 1994, \aap, 292, 381
\reference{Dej96} Dejonghe, H., De Bruyne, V., Vauterin, P., \& Zeilinger,
	W. W. 1996, \aap, 306, 363.
\reference{DdZ88} Dejonghe, H. \& de Zeeuw, P. T. 1988, \apj, 343, 113
\reference{ZLB85} de Zeeuw, P. T. \& Lynden-Bell, D. 1985, \mnras, 215, 713
\reference{DSW98} Dutta, S. N., Statler, T. S., \& Weil, M. 1998, in
	preparation
\reference{Fab97} Faber, S. M., Tremaine, S., Ajhar, E. A., Byun, Y.-I.,
	Dressler, A., Gebhardt, K., Grillmair, C., Kormendy, J., Lauer, T. R.,
	\& Richstone, D. 1997, \aj, in press
\reference{Fas96} Fasano, G. 1996, in Fresh Views of Elliptical
	Galaxies, A. Buzzoni, A. Renzini, \& A. Serrano, eds.
	(San Francisco: Astronomical Society of the Pacific), p. 37
\reference{FIH89} Franx, M., Illingworth, G. D., \& Heckman, T. 1989,
	\aj, 98, 538
\reference{FIZ91} Franx, M., Illingworth, G. \& de Zeeuw, T. 1991, \apj,
	383, 112
\reference{KoB96} Kormendy, J. \& Bender, R. 1996, \apjl, 464, L119
\reference{Lau95} Lauer, T. R., Ajhar, E. A., Byun, Y.-I., Dressler, A.,
	Faber, S. M., Grillmair, C., Kormendy, J., Richstone, D., \& Tremaine,
	S. 1995, \aj, 110, 2622
\reference{MaB94} Magorrian, J. \& Binney, J. 1994, \mnras, 271, 949
\reference{Mer97} Merritt, D. 1997, \apj, 486, 102
\reference{MeF96} Merritt, D. \& Fridman, T. 1996, \apj, 460, 136
\reference{MeV96} Merritt, D. \& Valluri, M. 1996, \apj, 471, 82
\reference{Nie91} Nieto, J.-L., Bender, R., \& Surma, P. 1991 \aap, 244, L37
\reference{Rix97} Rix, H.-W., de Zeeuw, P. T., Cretton, N., van der Marel,
	R. P., \& Carollo, C. M. 1997, \apj, 488, 702
\reference{Sat80} Satoh, C. 1980 \pasj, 32, 41
\reference{Sta94a} Statler, T. S. 1994a, \apj, 425, 458 (S94a)
\reference{Sta94b} Statler, T. S. 1994b, \apj, 425, 500 (S94b)
\reference{Sta94c} Statler, T. S. 1994c, \aj, 108, 111 (S94c)
\reference{StF94} Statler, T. S. \& Fry, A. M. 1994, \apj, 425, 481
\reference{SSC} Statler, T. S., Smecker-Hane, T., \& Cecil, G., \aj, 111,
	1512 (SSC)
\reference{MBD90} van der Marel, R. P., Binney, J. J., \& Davies, R. L. 1990,
	\mnras, 245, 582 (MBD)
\reference{vdM91} van der Marel, R. P. 1991, \mnras, 253, 710
\reference{Whi97} Whitmore, B. C., Miller, B. W., Schweizer, F., \& Fall,
	S. M. 1997, \aj, 114, 1797

\end{references}
\end{document}